# Conceptual Modeling of a Procurement Process:

Case study of RFP for Public Key Infrastructure

Sabah Al-Fedaghi
Computer Engineering Department
Kuwait University
Kuwait
sabah.alfedaghi@ku.edu.kw

Mona Al-Otaibi
Information Technology Department
Ministry of Finance
Kuwait
malotaibi@mof.gov.kw

*Abstract*—**Procurement refers to a process resulting in delivery of goods or services within a set time period. The process includes aspects of purchasing, specifications to be met, and solicitation notifications as in the case of Request For Proposals (RFPs). Typically such an RFP is described in a verbal ad hoc fashion, in English, with tables and graphs, resulting in imprecise specifications of requirements. It has been proposed that BPMN diagrams be used to specify requirements to be included in RFP. This paper is a merger of three topics: (i) Procurement development with a focus on operational specification of RFP, (ii) Public key infrastructure (PKI) as an RFP subject, and (iii) Conceptual modeling that produces a diagram as a supplement to an RFP to clarify requirements more precisely than traditional tools such as natural language, tables, and ad hoc graphs.**

## I. Introduction

*Procurement* refers to "a careful, usually documented process resulting in delivery of goods or services within a set time period" [1]. In project management the process includes aspects of purchasing, specifications to be met, and solicitation notifications. Procurement, also known as purchasing and supply, "is amongst the key links in the supply chain and as such can have a significant influence on the overall success of the organization" [2]. Without loss of generality the present study focuses on the first phase of the procurement process, which includes needs specification and construction of the request for proposal (RFP).

### A. Problem and solutions

Typically an RFP is described in a verbal ad hoc fashion, in English, with tables and graphs, resulting in imprecise specifications of requirements. Challenges of the traditional RFP approach include difficulty in holding vendors accountable, and contract management issues that often result in massive change requests and overruns [3].

Organizations that are in the process of developing a Request for Proposal (RFP) have often looked to existing sources for ideas on *how to phrase language* to cover a specific topic. They are often disappointed to learn that the search for RFP language examples is a time-consuming exercise that involves searching across multiple publications that may or may not include the topical information that they seek. (Italics added)

According to [4], it is quite common to see RFPs with requirements that are very broad, derived from a vendor's list of features, or copied from another organization's RFP. Among their suggested remedies is to prepare diagrams of the RFP process. "Model your business process graphically. Business process diagrams (or models) are excellent at showing gaps in the process or errors in your understanding" [4]. They particularly recommend Swim Lane diagrams.

Hadrian and Evequoz [5] enumerate the main difficulties in RFP requirements specification:

- Expressing precisely what will be needed (i.e., specific requirements and attaching requirements to specific parts in a process).
- Expressing requirements in a standardized form.
- Tracing requirements coming from different sources

In general, according to Hadrian and Evequoz [5], a methodology to produce more precise requirement specifications would be helpful for all stakeholders. Requirements should be unambiguous and validated by business users. Hadrian and Evequoz [5] proposed use of BPMN diagrams [6] to specify requirements to be included in Request for Proposals. BPMN is an International standard for process documentation that bridges the gap between business and IT people.

Similarly, we propose applying a conceptual model (the Flowthing Machine, FM) that can be used to facilitate creation of RFP specifications. This can then be used by all stakeholders in the process, since FM is a conceptual model that can be understood without substantial knowledge of technical details. Hence, the aim in the next section is to demonstrate that FM can be utilized as a tool for a comprehensive expression of what is needed. It is understood that, initially, developing an RFP entails a certain amount of guesswork about details. An advantage of FM is that the drawing can be modified fairly easily as details evolve.

### B. Additional problem: Communication among stakeholders

An additional problem in requirements specification for an RFP is related to communication among stakeholders. In a government RFP [7], it is stated that,

The assumptions, assessments, statements and information contained in this RFP, may not be complete, accurate,

adequate or correct. Each Bidder should, therefore, conduct their own investigations and analysis and should check the accuracy, reliability and completeness of assumptions, assessments and information contained in this RFP and obtain independent advice from appropriate sources.

A general aim of this paper is to introduce a modeling language that expresses the technical parts of the RFP in a "neutral" representation that facilitates communication among stakeholders.

Public key infrastructure (PKI) is intentionally selected as the content of RFP because "all of the books or Web sites on the subject either assume that you already know all about PKI or they use so many big words that they are hard for a beginner to understand" [8]. PKI is suitable as a test case for communication among stakeholders by providing a non-technical language that underlies the RFP.

A *neutral* (i.e., independent of whatever technology is used) *representation*, mentioned previously, is a product of the FM conceptual model. This paper considers the topic of conceptual *modeling* in order to demonstrate its advantages in the field of software engineering for procurements. Consequently, this paper is a merger of three areas of study:
1. Procurement development with a focus on operational specification of RFP
2. Public key infrastructure (PKI) as an RFP subject
3. Conceptual modeling that produces a diagrammatic description as a supplement to the RFP for clarifying requirements in a more precise manner than traditional tools such as natural language, tables, and ad hoc graphs.

*C. Conceptual modeling*

Twenty years ago, modeling of systems was viewed as a great discovery for accelerating resolution to challenges to manufacturing industries by 2020 [9]. One major scientific area that embraces modeling is software engineering. Software is everywhere in the infrastructure and affects all fields of life. Software engineers deal with more complex problems than any other engineering discipline [10]. Decades of work on software abstraction have helped gain intellectual control over systems of ever-increasing complexity. This has motivated adopting a modeling approach throughout the software development process with tools such as UML and SysML.

According to Armstrong [3], the traditional RFP process involves a phased approach similar to a waterfall: a requirements specification phase, system requirements phases, a design phase, and an implementation phase. Requirements specification is a basic phase in software lifecycle system development. Software engineers have put much effort into the process of transforming requirements into software architecture, including creating a *text description* of the envisioned system as well as creating *models*. The key problem is to give an unambiguous, easy to understand description of a system and how it works. "We can do so with English descriptions; but such descriptions are often cumbersome, incomplete, ambiguous and can lead to misunderstandings" [11].

Armstrong [3] recommended incorporating Agile into an adaptive collaborative development process, *significantly leveraging UML for modeling*, using a comprehensive traceability strategy, and automatically generating RFPs. In the first iteration of the process, "a business *use case model* that include[s] coarse-grained business workflow diagrams (activity diagrams) [and] business use case outlines" [3]. Later the process would incorporate development of UML collaboration diagrams for business use cases, and class diagrams for business participant responsibilities.

Douraid et al. [2] modeled the procurement process at the operational level by using UML to describe the static and dynamic behavior of the system [12]. "UML is not restricted to modeling software. It is also used for business process modeling, systems engineering modeling and representing organizational structures. It is a general-purpose modeling language that includes a graphical notation used to create an abstract model of a system. It is designed to specify, visualize, construct, and document software-intensive systems" [2].

*D. Approach*

The aim of this paper is to supplement the RFP with a model, i.e., diagrams that express how the features and services of PKI would logically operate in the proposed system. Such an approach is not new, and the following is an illustrative example.

In requests for proposals by the Judicial Council of California [13], proposers must respond to "Use Case Scenarios with a narrative response describing how their product features and or services will excel or be challenged in addressing these use case scenarios." An example (supported by a diagram) of such a use case is as follows:

A person, business or government agency brings a document to the clerk's office. The clerk records the document in the Case Management System (CMS) and receives a case number from the CMS (either for an existing case or as a newly filed case). A cover sheet is produced that contains the information that will be used as index values for this document. The cover sheet and document will be scanned into the Document Management System… .

The authors [13] provide a sample diagram of the PKI process accompanying an RFP showing how the agency conceives the workings of the PKI system. This does not impose a rigid method; rather it is an initial "solution" to the problem that the agency tries to solve; and the bidder can respond with a counter model that is a modification or replacement of this conceptualization (see Fig. 1).

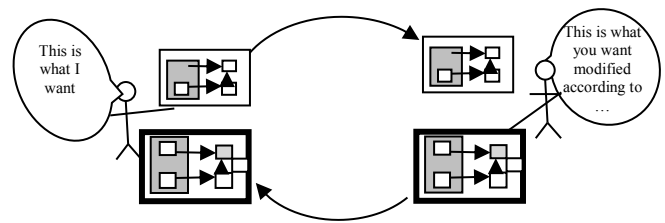

Fig. 1. Diagram showing how the system works

## II. FLOWTHING MACHINE (FM)

This section briefly reviews the FM model that forms the foundation of the theoretical development in this paper; however, the example given here is a new contribution.

### A. Basic notions

The FM model (see [14–23]) is a diagrammatic schema that uses *flowthings* (hereafter, *things*), defined as *what can be created, released, transferred, received*, and *processed*, by means of stages in a flow machine (Fig. 2). *Things* begin to flow through the stages of the machine when they are created by the machine or imported from other machines.

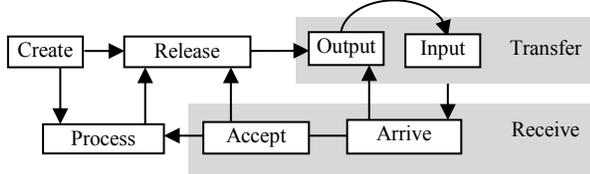

Fig. 2. Flow machine

*Flow* here entails transition or realization of change as well as movement and positioning. **Create** is the emergence of a thing in the system from outside it. The rest of the flow is succession from one stage to the next. Such flows are specified in analogy to drawing traffic flows on a city map. There, as will be discussed later, *dynamic* flows are shown in terms of *events* that describe the behavior of the system, when the streets of the city become streams of flow of cars, pedestrians, etc.

The point here is that the *flow* is often thought of as physical movement, but in FM, it can be much more than that. It is a notion that captures the *conceptual* movement of thought, sensation, being, and doing. The modeler builds a conceptual construct and also a conceptual "movement"; we call it flow. Thus, a physical house *flows* from one sphere (e.g., a *class* in UML terminology) to another when there is a change in ownership from a person to a certain bank, and a car on an assembly line *flows* to robots and workers *simultaneously* when it is processed, e.g., one fixes glass while another puts on tires, etc. Flows might be fast or slow, parallel or sequential, physical or digital (e.g., uploading software) or mental (e.g., inspecting finished products), or comprise only creating, only processing, etc.

The stages in Fig. 2 can be described as follows:
**Arrive**: A thing reaches a new machine.
**Accept**: A thing is approved to enter a machine. If arriving things are always accepted, *Arrive* and *Accept* can be combined as a **Receive** stage.
**Process** (change): The *thing* goes through some kind of transformation that changes its "state" without creating a new thing.
**Release**: A thing is marked as ready to be transferred outside the machine. Note that things can be released from a given system without being transferred, as in the case of sent emails waiting for a damaged channel to be fixed.
**Transfer**: The thing is transported somewhere from or to outside the machine.
**Create**: A new thing is born (created) in a machine.

Flow machines use the notions of *spheres and subspheres*. These are constructs (mental conceptions) of machines and submachines. Multiple machines can exist in a sphere if needed. A sphere can be a person, an organ, an entity (e.g., a company, a customer), a location (a laboratory, a waiting room), a communication medium (a channel, a wire). A machine is a subsphere that embodies the flow; it itself has no subspheres. This sphere notion is taken from cognitive linguistics where an *idea* is treated as complex units associated with other entities or other forms of association. "A door, for example, also connotes a door knob, a key hole, a door jamb, etc." [17].

FM also utilizes the notion of *triggering*. Triggering is the activation of a flow, denoted in the machine diagrams by a *dashed arrow*. It is a dependency relationship among flows and parts of flows. A flow is said to be triggered if it is created or activated by another flow (e.g., a flow of electricity triggers a flow of heat), or activated by another point in the flow. Triggering can also be used to initiate events such as starting up a machine (e.g., by remote signal). Multiple machines can interact by triggering events related to other machines in those machines' spheres and stages.

### B. Example

Douraid et al. [2] introduced a model for generally depicting a procurement process, including supplier management, inventory management, and invoicing and delivery procedures. Their set of conceptual and UML models was designed for use in constructing a simulation framework for a procurement process. "The behavioral aspect is captured from activity and state diagrams to characterize the dynamic side of our approach" [2]. Figs. 3 and 4 show partial views of their state and activity diagrams.

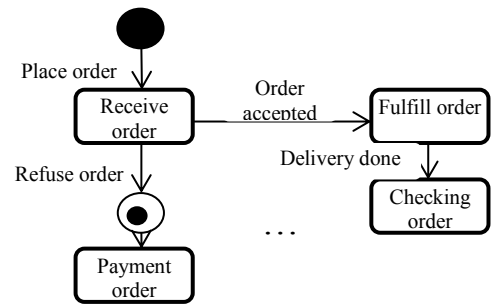

Fig. 3. Order state diagram (redrawn, partial from [2])

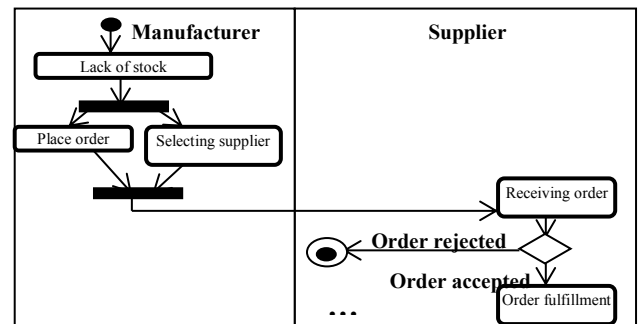

Fig. 4. Supplier-manufacturer relationship activity diagram (redrawn, partial from [2])

Fig. 5 shows the corresponding FM representation of this supplier-manufacturer relationship. First, the storage of the manufacturer (circle 1 in the figure) is processed (checked), and if there is a lack of stock (2) then this *triggers*,

- Generating data, e.g., item name, quantity (3), and
- Selecting a supplier (4)

Accordingly, these two *things* flow to an ordering management procedure (5) that triggers the creation of an order (6).

The order flows to the supplier (8) where it is processed.
- If the order is rejected, a negative response is sent back (9).
- If the order is accepted, a positive response is sent (10). Additionally, the goods are released (11) and sent to the manufacturer (12).

There the goods are processed (13),
- If acceptable, (14) they are sent to storage (15). Additionally, a payment is made (16) and sent to the supplier (17).

If the goods are not acceptable (18), they are returned to the supplier (19).

### III. CASE STUDY: PUBLIC KEY INFRASTRUCTURE

The aim of eGovernance is to automate government operations, business processes, and service delivery online. As a result, electronic documents are infiltrating every aspect of the government workflow. Difficulties arise when a signature authorization is needed that requires a physical signature.

This manual process increases costs and time, and impede the benefits of a fully electronic workflow. Digital Signatures provide a solution for creating legally enforceable electronic records while eliminating the need to print documents for signing.

A digital signature can be used to authenticate the identity of the sender of a message or the signer of a document. Here we assume general knowledge of public key cryptography since a digital signature requires a key pair: the *Public* and *Private* Keys.

The private key is retained by the owner and the public key is public for everyone. Information encrypted by a private key can be decrypted only by means of the corresponding public key. Because of our case study, this paper focuses on certificate authorities (CAs) instead of such approaches as web of trust and simple public key infrastructure.

*Public Key Infrastructures* is a support system for usage of public key cryptography [24]. It includes all hardware, software, people, policies, and procedures for creating and handling digital certificates and manages public-key encryption. This is accomplished through (i) providing digital signatures with (ii) verification of the ownership of public keys. Common PKI functions include issuing certificates, revoking certificates, storing and retrieving certificates. Enhanced functions include time-stamping and policy-based certificate validation.

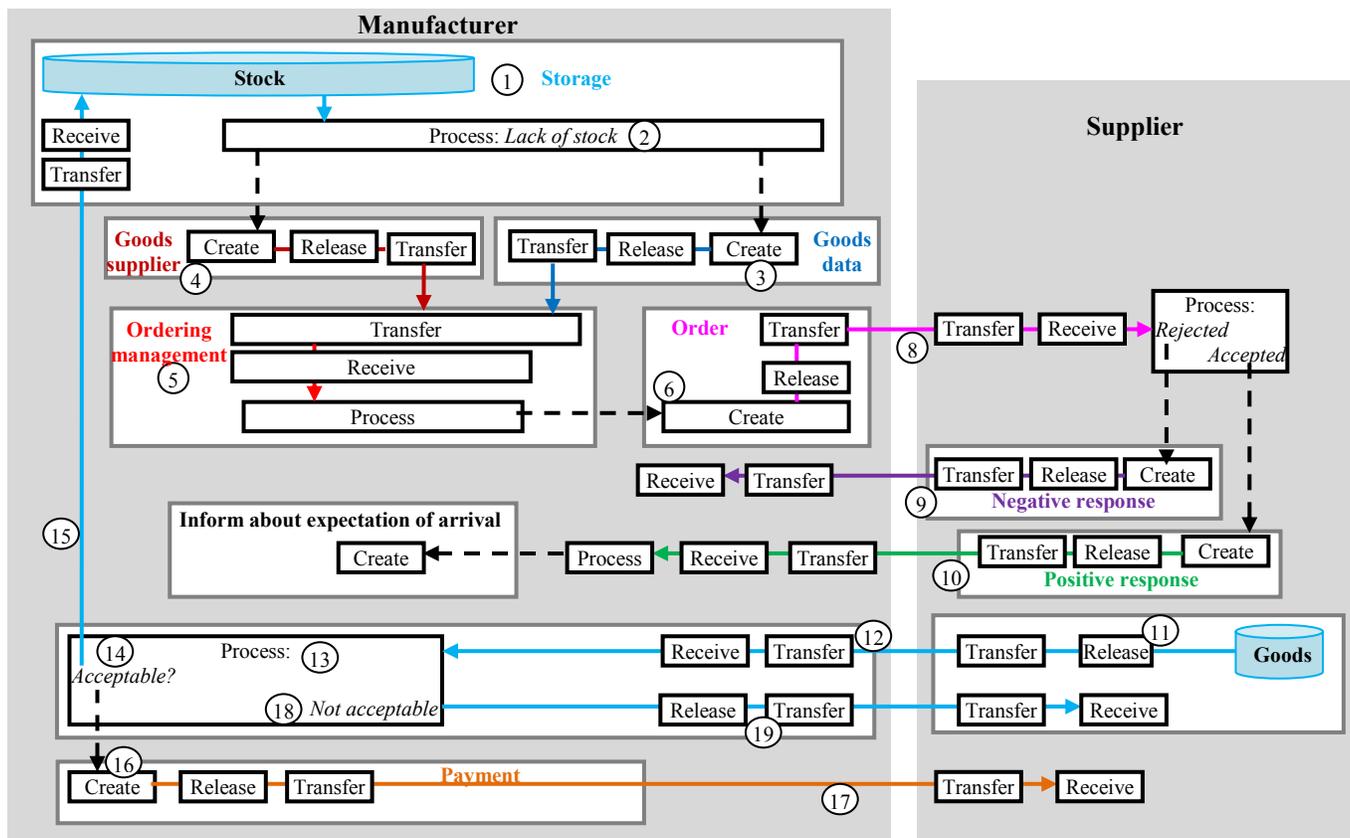

Fig. 5. FM representation of the example

## A. How to create a digital signature

In a digital signature, a process called "hashing" converts the data to what is called a *message digest* which is encrypted with the *private key* to produce the digital signature that is appended to a document.

**Example** (from [8]): Suppose that *I* need to send *you* an e-mail message. Assume that the message does not need to be encrypted, but that what is needed is as follows (see Fig. 6):
- *Assurance that the message came from **me***.
- *Verification that the message was not intercepted and altered in transit.*

Assume that the message is *The check is in the mail*.

1- *I* produce a non-reversible hash of the message. That is, I create a hash by adding together the ASCII values of each character in the message: 84 + 104 + 101 + 32 + 99 + 104 + 101 + 99 + 107 + 32 + 105 + 115 + 32 + 105 + 110 + 32 + 116 + 104 + 101 + 32 + 109 + 97 + 105 + 108 + 46 = Â 2180. The hash 2180 is non-reversible because there is no way that we produce from 2180 the message *The check is in the mail*.
2- The hash is appended to the end of the message *The check is in the mail*.
3- *I* use my private key to encrypt the hash value 2180 and append it to the end of the message before I transmit it to you.
4- When *you* receive the message, you calculate the message's hash by using the same algorithm that was used to produce the hash in the first place. If *you* calculate the same value as the hash value that is appended to the end of the message, then you can be sure that *the message has not been altered in transit*.
5- *You* use my public key to decrypt the hash value. If you are able to do this successfully, then you know beyond doubt that *I am the one who encrypted the hash value*.

## B. Certificates

A PKI is based on things called *certificates* that are issued by the Certificate Authority (CA) and serve as digital identification. Certificates associate users with their public keys. They can be created by way of software, and we limit our interest in this paper to a standard that defines the format of public key certificates required in the *case study* that will be discussed later.

We assume here that the CA generates the public and private keys for the user. The public key has to be signed by the CA, where:
1. The CA uses a hash algorithm to generate the so-called *digest*,
2. The digest is encrypted with a private key. The result is a digital signature.
3. The CA then makes the digitally signed certificate available for download to the person who requested it.

In general the Public Key Infrastructure works as follows:

> A user applies for a certificate with his public key at a registration authority (RA). The latter confirms the user's identity to the certification authority (CA) which in turn issues the certificate. The user can then digitally sign a contract using his new certificate. His identity is then checked by the contracting party with a validation authority (VA) which again receives information about issued certificates by the certification authority. [25]

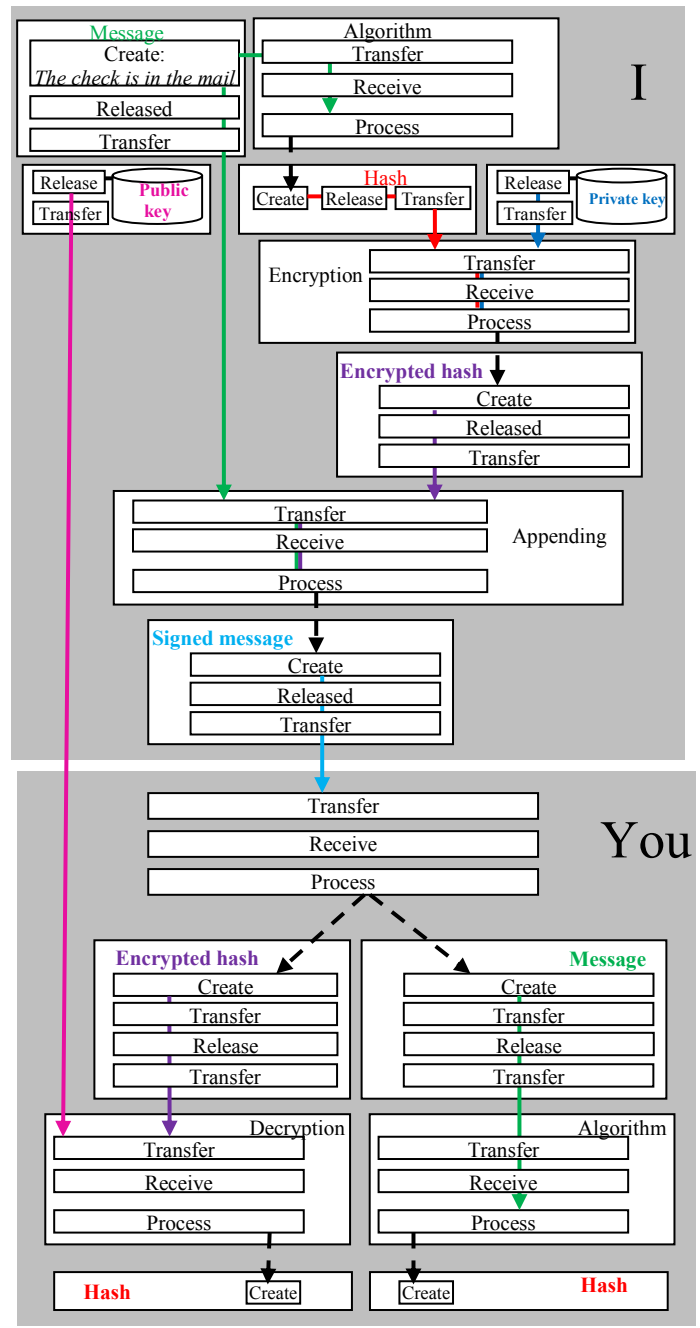

Fig. 6. Example that illustrates a digital signature

## IV. RFP CASE STUDY

The case study discussed in this section involves a government agency that seeks the services of a bidder specialized in Enterprise Public Key Infrastructure (PKI) services.

## A. General description of the RFP

The RFP contains 59 pages, including a section on the Current Environment with a general view of existing infrastructure, mainframe, and network base IT infrastructure. Of interest in this paper is the section where CA/RA functional and technical requirements are described. In the RFP, the section titled *Certificate Issuance and PKI Lifecycle Management* is a mix of textual description and diagrams. The diagrams are mostly textbook illustrations such as the one shown in Fig.7.

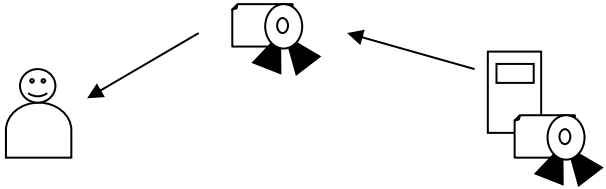

Fig. 7. Example of a diagram used in the RFP (redrawn)

A sample text is the following.

**Certificate Authority**
 The key generation and certification services must be used with a Registration Authority (RA) Server. The CA Server is a PKI Server including:
- Consists of CAs with their own certificate signing keys and other parameters from one Server instance
- Provides simplified server-side key generation and client-side key generation
- Provides RSA certificate signing with keys of 1024, 2048, 4096 bits

**Certificate Validation**
 Proposed OCSP Server must have an advanced x.509 certificate Validation Authority server that fully conforms to the IETF RFC 6960 standard. It is approved for use by US federal agencies for HSPD-12 implementations.

## B. A justification for incorporating FM as a supplement to the RFP

Even though it is clear that the main objective of the project is "to *identify* and implement the most appropriate PKI solution that fulfills the [Agency's] *requirements* to improve the security, accuracy, and agility of its IT Infrastructure," it is unclear what these *requirements* are. We will focus here on parts that describe digital signatures. Searching *all* instances of "signature" in the RFP, we copied the following requirements directly from the RFP text:

- Requesting and embedding timestamp responses, requesting and, requesting and embedding OCSP responses, PDF permissions, and server-side archiving of signed documents to disk.
- Creating own PKI systems for **Digital Signature** issuance and Staff logical access Smart card.
- Signing Server should be complete solution for creating and verifying **digital signatures** on document, web form or transaction.
- Server must provide autonomous and irrefutable proof of time for transactions, documents and **digital signatures**.
- Prove when a **digital signature** was applied by the signer so that its validity can be verified in the long-term, even after revocation of signer's digital credentials.
- PKI can provide robust user authentication and strong **digital signatures**.
- The USB should include **digital signatures** and encryption.
- Signing Server can create and verify all common **signature** formats.
- A **signature** service should have the flexibility to be integrated with any application either on the web or a local workstations. It should easily integrate the signing process into the business workflow.
- **Signature** services should be made obtainable for multiple devices and scenarios. It should work on the principle of 'Anytime, Anywhere, Any device' access. The signature capability should be integrated with client applications to allow for documents, emails, data, etc., to be easily signed by their intended signatories.
- **Signature** service should support What You See Is What You Sign (WYSIWYS).
- PDF and Document **signature** should provide visible signatures.

We point out the crucial role of Requirements Specifications within an RFP as the main basis for evaluation by bidders and for the challenges associated with gathering and specifying requirements. In general, according to Hadrian and Evequoz [5], while the legal basis that governs public procurements gives precise guidelines, there is a lack of clear instructions regarding the form and necessary content of a request for proposal.

## V. FM DESCRIPTION OF PUBLIC KEY INFRASTRUCTURE

This section includes a conceptual model of how the required system registers users, issues PKI certificates, and is used by the employees of the agency. It includes conceptual components that include hardware (e.g., servers), software, and manual operations.

### A. Issuing of certificates

Fig. 8 shows the FM representation of digital signature and certificate issuing under the PKI framework.

*Application for certificate*
 An employee (circle 1) chooses the option (2) to request a digital certificate through his/her account. The request flows (3) to the web interface server dedicated to the PKI system, then to (4) the server of the cryptographic service provider (5). The request process (5) triggers creation of the key (6), including a public key (7) and a private key (8).

*Registration Authority (RA)*
 An RA verifies the identity of employees requesting their digital certificates to be stored at the CA. RA functions include the processes of collecting user data and verifying user identity, which is then used to register a user.
 Accordingly, the created key flows to the server of RA (9) to be processed to stamp it with a validation period (10) and to verify the employee's identity.

*Certificate Authority*
 Then, the RA passes the keys with their validation information to the Certificate Authority (CA) system (11). The CA combines validation data (12) and the public key (14) with other information (identity proof, name of CA, and serial number) to create the Digital Key Certificate (15). The private key (13) is kept separately for later encryption of signed documents.
 Accordingly, the digital key certificate (15) and the private key (13) are stored in the Database (Repository) (16) to be ready for the employee's use. The database is a secure location in which to store and index keys. An acknowledge-1 is sent to the employee to inform about creating and storing the digital certificate. The acknowledge-1 instructs the employee on the next step, which is to request (18) digital signature creation (19).

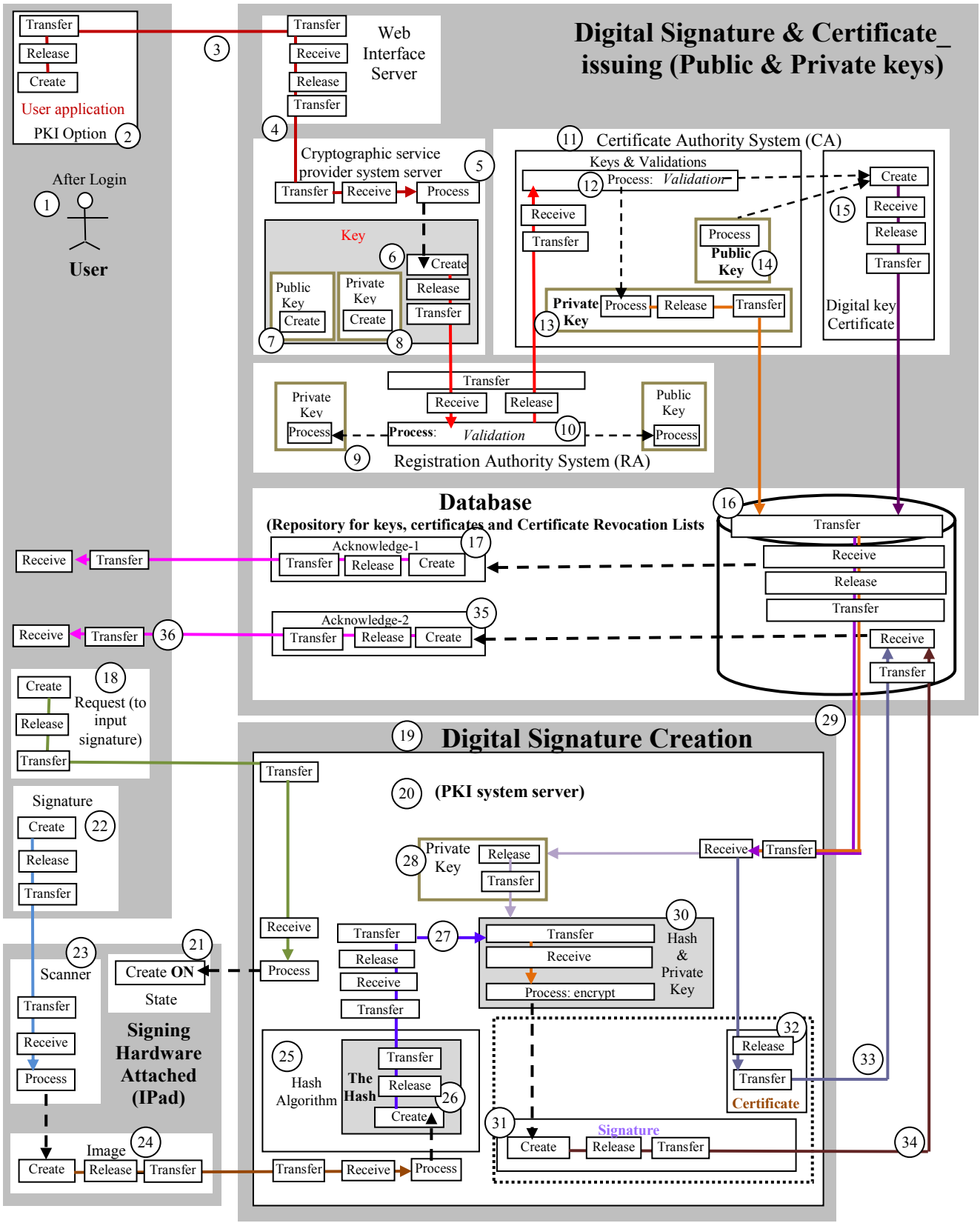

Fig. 8. FM description of the digital certificate as conceptualized by the agency (extracted from the RFP and general knowledge of the subject.

*Digital Signature Creation*

The digital signature request is received (20), triggering turning ON the Signing Hardware Attached (iPad) (21) to enable the employee to input his/her signature (22) through the scanner (23). The scanner (23) sends the image (24) of the signature to the PKI system server (20). The image is hashed using a special hash algorithm (25). The created hash (26) flows (27) to be combined with the private key (28) which was sent (29) earlier. The hash and private key (30) are encrypted to trigger the signature (31) along with the digital key certificate (32) to flow together (33 to the database (repository) (16) to be stored, producing an acknowledge-2 (35) that flows to the employee (36).

Fig. 8 provides a basis for communication and explanation. Multilevel simplifications of the figure can be made for different purposes such as presentations for high-level technical management. For example, Fig. 9 shows the figure simplified after all depiction of stages has been omitted.

B. *Digitally signed document*

Signing a document digitally is modeled in Fig. 10. A user (circle 1) selects to request (2) signing a document (3) which is already stored on the user's computer. The document flows (4) to the PKI system (5) to be processed using a hash algorithm (6). The created hash (7) flows (8) to be combined with the private key (9) in the CA repository.

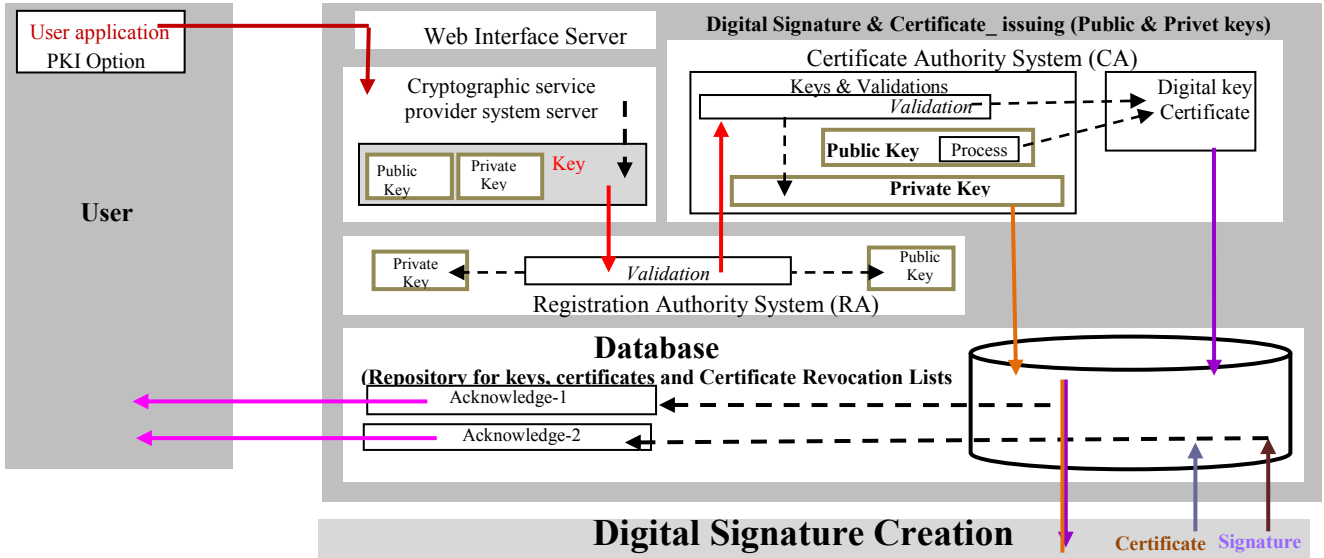

Fig. 9. Simplification of the upper part of Fig. 8 by deletion of stages

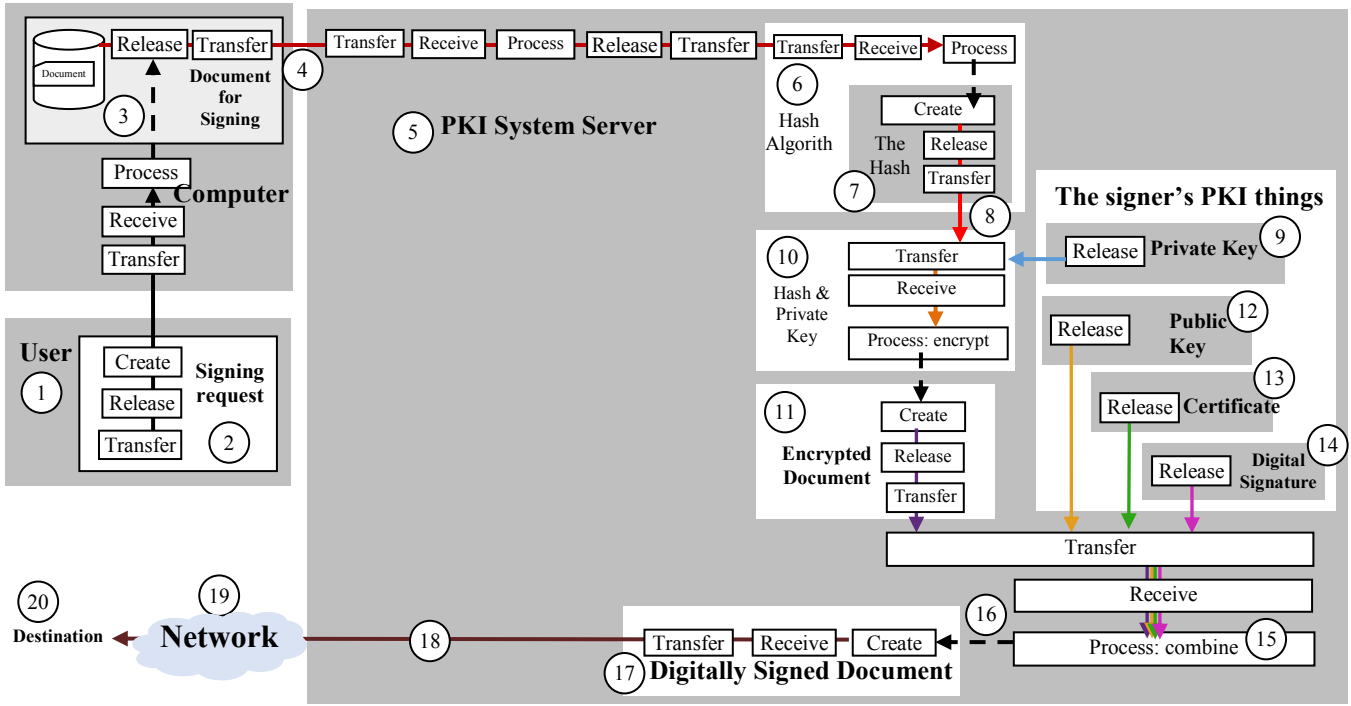

Fig. 10. FM description of the process of digitally signing a document.

The hash & private key (10) are encrypted to create an encrypted document (11). Then, the encrypted document (11) is combined (15) with the other signer's PKI Objects (the public key (12), the certificate (13), and the signature (14)) to create the Digitally Signed Document (16). The digitally signed document (17) is sent (18) through the network (19) to its destination (20).

*C. Decrypting the received document*

As shown in Fig. 11, a user (Recipient) (circle 1) selects to request (2) decrypting a received digitally signed document (3) that is already loaded on the recipient's computer. The document (3) flows (4) to the PKI system (5) to be processed (6). Processing separates the encrypted document (7) from the signer's PKI certificate (10), which contains the public key (8) and the digital signature (9). Using the public key (8), two decrypt operations (11 and 12) are applied to the encrypted document (7) and the digital signature (9). Decryption (11) triggers creating the document (13) to be hashed (14) in order to create the hash (15), additionally decryption (12) triggers creating the hash (16). The two hashes (15 and 16) are compared (17) (equal or not) to verify the sender's identity and validate his or her signature.

*D. Additional general specifications*

General specifications can be superimposed (in their correct places) on the FM diagrams, including
- (CA) Server specifications such as using a web services interface like XML/SOAP.
- Supporting of X.509 standard.
- Providing RSA certificate signing with, say, 4096 bits
- Supporting several hash algorithms, e.g., SHA-1, SHA-2

The diagrams can also be expanded to include:
- Backup
- Time Stamp Authority

## VI. CONCLUSION

This paper has introduced a diagrammatic conceptual representation (FM) as a tool for the specification of requirements in RFPs. The FM model includes basic elements of *things*, their *flows*, and their *stages*, within *spheres* that overlap with other *spheres*. FM is applied to a sample case study of RFP for public key infrastructure (PKI). The results indicate the following:
1. FM is viable as a modeling tool that complements RFP.
2. FM lends itself as a theoretical base for defining requirements in procurements.

The complex FM diagrams may present difficulties; however, some solutions to visual complexity have already been already been implemented in many engineering systems (e.g., aircraft and high-rise building schemata) through multilevel simplifications, as we did in Fig. 9. The details can be lumped together by omitting stages and unifying flows in the model. Nevertheless, the underlying FM schema remains the reference for any further usage such as analysis and documentation.

Further research will work on other types of RFPs. Many issues remain to be clarified; however, this paper demonstrates potential feasibility of the approach.

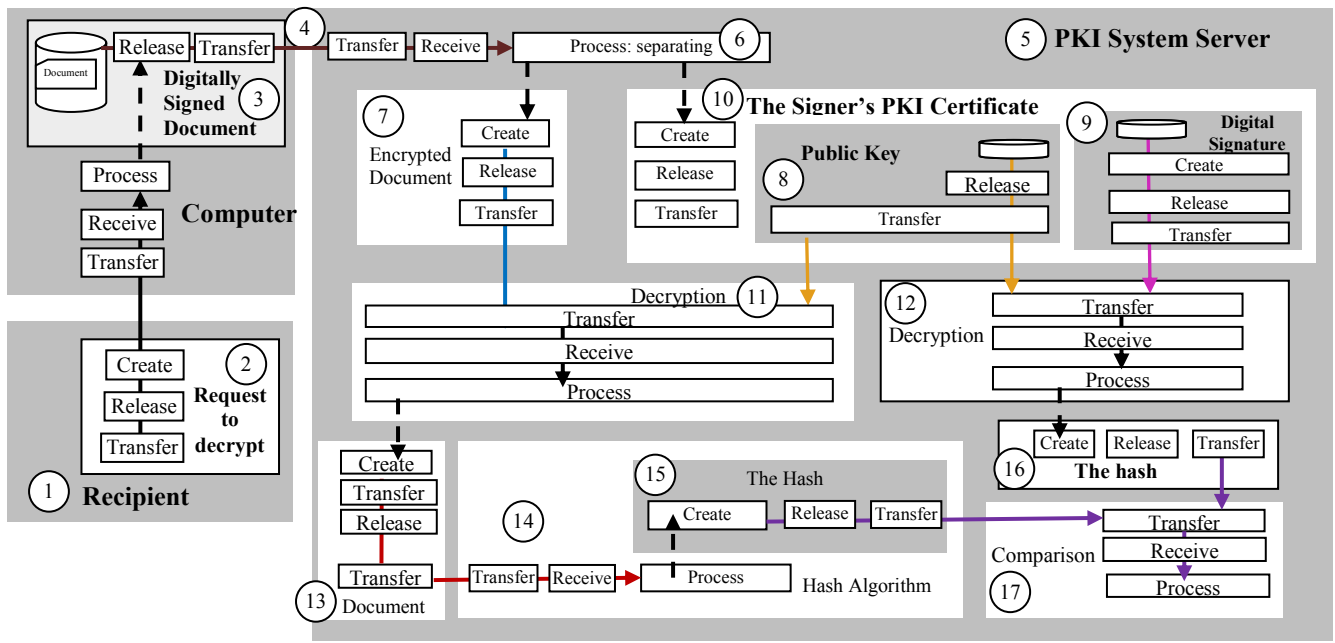

Fig. 11. FM description of the process of decrypting a document.